\documentclass[aps,prb,showpacs,twocolumn,superscriptaddress]{revtex4}

\usepackage{graphicx}
\usepackage{amsmath}
\usepackage{natbib}
\usepackage{bm}
\usepackage[dvips]{color}

\begin{document}

\title{Effects of electronic correlation on X-Ray absorption and dichroic spectra at L$_{2,3}$ edge}
\date{\today}

\author{L.~Pardini, V.~Bellini and F.~Manghi}
\email[Electronic address: ]{franca.manghi@unimore.it}
\affiliation{CNR - Institute of nanoSciences - S3, Via Campi 213/A, I-41125 Modena, Italy  \\ and Dipartimento di Fisica,
Universit\`a di Modena e Reggio Emilia, Via Campi 213/A, I-41125
Modena, Italy}

\begin{abstract}

We present a new theoretical approach to describe X-Ray absorption
and Magnetic Circular Dichroism spectra in the presence of e-e
correlation. Our approach   provides an unified picture to include
correlations  in both charged and neutral excitations, namely in
direct / inversion photoemission where electrons are removed/added,
and photo absorption where electrons are promoted from core levels
to empty states. We apply this approach to the prototypical case of
L$_{2,3}$ edge of 3$d$ transition metals and we show that the
inclusion of many body effects in the core level excitations is
essential to reproduce, together with satellite structures in core
level photoemission, the observed asymmetric line shapes in X-ray
absorption and dichroic spectra.

\end{abstract}
 \maketitle

The excitation due to X-ray absorption  is a complex
phenomenon where many body effects may play a major role. The
independent particle picture has been shown to be inadequate in
describing features of X-ray absorption spectra in transition metals
\cite{Ankudinov} and their compounds \cite{kotani2,kotliar} where
e-e interactions are known to be non negligible. To account for
these failures various theoretical methods have been developed that
differ on how the e-e Coulomb interaction is taken into account.
They range from atomistic approaches, where a parameterized many body
Hamiltonian is solved via configuration interaction methods
\cite{kotani2}, to  solid state first-principles schemes that include
properly the structure of valence electrons
\cite{Soininen,Ambrosch-Draxl} but loose some of the atomic
many-body effects that can be relevant to the physics of the
process.

In this paper we present a new approach that treats on the same
footing the localized and itinerant character of electrons in a solid,
exploiting on one side the first principle calculations of the band
structure and on the other side the localized picture to treat many
body effects associated to electron-electron interactions.
We will show that the on-site interaction between core and valence
electrons, responsible of satellite structures in core level
photoemission spectra, gives rise to  asymmetric line shapes in XAS
and XMCD spectra in agreement with experiments. Application to the
prototypical case of absorption from the $L_{2,3}$ edge  in 3$d$
transition metals is presented.

Within the  independent-particle scheme the X-ray absorption is
described as the addition of one electron-hole pair to a
non-interacting Fermi sea. In the language of many-body theory, this
corresponds to approximate the two-particle polarization propagator
to zeroth order. It is possible to improve this approach by
substituting the bare particle and hole propagators with the dressed
ones. Except for vertex corrections this would be the {\emph exact}
two-particle propagator. This picture is very physical in the sense
that the creation of the e-h pair due to X-ray absorption can be
visualized as the removal of one core electron  in the presence of
the valence continuum  and its addition to the conduction band, a
sort of superposition of photoemission and inverse photoemission
spectra, plus possible electron-hole interaction.

Even neglecting the excitonic effects associated to the
electron-hole interaction we are left with  a very hard task,
namely the calculation of hole and particle propagators of an
interacting many particle system. This is even more difficult here
since we are dealing with two processes that from the point of view
of e-e interactions are very different: on one side the removal of
one electron from an inner core state with atomistic interactions
resulting in multiplet structures that in most cases survive the
band formation; on the other side the addition of one electron in an
itinerant conduction state where local short range e-e interactions
coexist with hopping from site to site. An approach that can treat
these two apparently opposite situations in an unique way is based
on the generalized Hubbard model where the different strength of
localization can be taken into account in terms of different $U$ and $J$
parameters and different band widths (finite and zero band width for
valence states and non-dispersive inner core energy levels
respectively).  In the present case the Hubbard Hamiltonian can be
usefully partitioned in core ($c$) and valence ($v$) contributions

\begin{equation}
\label{hh} \hat{H}= \hat{H}_{v}+ \hat{H}_{c}  .
 \end{equation}
The first term describes valence states in terms of single particle
band eigenvalues ($\epsilon_{ k n\sigma}$) and of on-site Coulomb
($U_{vv}$)   and   exchange  ($J_{vv}$) interaction
\[
\hat{H}_{v} = \sum_{k n \sigma}\epsilon_{kn \sigma} \hat{n}_{kn
\sigma} + \frac{1}{2}\sum_{i \sigma \sigma'}
(U_{vv}-J_{vv}\delta_{\sigma\sigma'}) \hat{n}_{i v \sigma}
\hat{n}_{i v \sigma'}
\]
Here $i$ is the site index.
Similarly
\[\hat{H}_{c}=\hat{H}_{cc}+\hat{H}_{cv}\]
where
\[
\hat{H}_{cc} = \sum_ {i \sigma} \left [ \epsilon_{c \sigma}
\hat{n}_{i c\sigma} + U_{cc} \hat{n}_{i c \sigma} \hat{n}_{i c -\sigma} \right ]\nonumber \\
\]
\begin{eqnarray}
\label{cc}
 \hat{H}_{cv} &=& \frac{1}{2}\sum_{kk'p}\sum_{\sigma
\sigma'} (U_{cv}-J_{cv}\delta_{\sigma \sigma'})  C^{n*}_{k'\sigma'}
C^n_{k'-p\sigma'}  \\ \nonumber &\times&
\hat{a}^{c\dagger}_{k\sigma} \hat{a}^c_{k+p\sigma}
\hat{a}^{n\dagger}_{k'\sigma'} \hat{a}^n_{k'-p \sigma'}.
\end{eqnarray}
Notice that $\hat{H}_{c}$ includes both core-core ($U_{cc}$) and
core-valence ($U_{cv}$,$J_{cv}$) e-e interaction. In the present
case we are interested in L$_{2,3}$ edge excitations, therefore we will
consider Coulomb and exchange integrals involving 2$p$ and 3$d$
orbitals.

The orbital character of the valence band comes into our scheme
through the  $d$ orbital coefficients $C^n_{\sigma}(k)$ of eq.
\ref{cc} obtained in our case  by ab-initio band structure
calculation. Both core and valence states have been calculated
within the full-potential, linearized plane-wave method (FLAPW) as
implemented in the \textsc{Wien2k} code \cite{wien2k}. Spin-orbit
coupling has been included in order to reproduce a non-zero total
orbital momentum that is one of the quantities that can be extracted
from dichroic spectra. For core states the relativistic effects are
accounted for by solving the Dirac equation while for valence states
the spin-orbit coupling is treated by perturbation theory
\cite{kunes2001b}.

The ab-initio band structure results could also be used  to estimate the U and J parameters: the bare on-site valence - valence   and core - valence  coulomb and exchange interactions can be explicitly calculated using the core and valence wavefunctions  and properly scaled   to account for screening effects. This procedure has been applied to the evaluation of core-core integrals in transition metals oxides assuming a static effective dielectric constant \cite{rozzi2005}.
Various ab-initio estimate of on-site  valence-valence Coulomb and exchange integrals have been recently proposed  for 3$d$ transition metals, either within a constrained density functional approach \cite{Cococcioni} or as dynamicaly screened on-site integrals \cite{Takashi}. The results depend on the procedure adopted and the range of variation in the parameters (1 to 5 eV) remains the same reported previously in the literature \cite{Steiner}.  In the present calculation U and J  have been used as adjustable parameters, with values for valence-valence within the above mentioned range, tuned  to optimize the agreement between theory and experiment. Their values are reported in table \ref{UJ}.  These values for the valence-valence interactions give an accurate description of the quasiparticle band dispersion probed by angular resolved photoemission \cite{jaime1,jaime2,Bellini}. The same is true of core-core interactions that have been fixed to reproduce the satellite position in core level photoemission spectra (see next figure on core level photoemission).

When dealing with charged excitations where one electron is removed
from a core level or added to an empty valence state one can solve
independently $\hat{H}_{c}$ or $\hat{H}_{v}$ to obtain hole and
particle spectral functions respectively. This is done here using
the 3-body scattering (3BS) approach that has been implemented to
treat  both core and valence spectra \cite{Monastra,Bellini,
rozzi2000, rozzi2005}: a charged excitation is described in terms of
scattering between the single particle state with one removed/added
particle and  the excited configurations containing one extra e-h
pair.

\begin{table}[t]
 \begin{center}
  \caption{\label{UJ} Coulomb and exchange integrals  (in eV)
describing the interaction between valence ($3d$) and core ($2p$)
states.  }
  \begin{ruledtabular}
   \begin{tabular}{lccccc}
                 & $U_{dd}$ & $J_{dd}$   & $U_{pd}$   &$J_{pd}$   \\
\hline
Fe               & 1.5      & 0.9            & 0.8       & 0.1       \\
Co               & 2.1      & 0.9            & 1.8       & 0.2       \\
Ni               & 2.0      & 0.9            & 2.5       & 0.2       \\
   \end{tabular}
  \end{ruledtabular}
 \end{center}
\end{table}

Energy and spin dependent complex self-energies
$\Sigma^{c/v}_{\sigma}(\omega)$ are calculated and from them
spectral functions for core  and valence states given by
\begin{equation}
\label{A}
 A^{c/v}_{\sigma}( \omega)=\frac{1}{\pi} Im
\frac{1}{\omega-\epsilon^{c/v}_{
\sigma}-\Sigma^{c/v}_{\sigma}(\omega)} \end{equation}
Please notice that for valence states the spectral function depends on band index and k-vector: $A^{v}_{\sigma}(\omega) \equiv A_{k n \sigma}(\omega)$ the k- and band index dependence coming from  $\epsilon^{v}\equiv \epsilon_{k n}$ and possibly from the self-energy.

Let us start from the core hole excitation. Within the present
approach  the hole self energy turns out to be \cite{Bellini,rozzi2000}
\begin{eqnarray}
\label{sigmac}
    \Sigma^c_{\sigma }(\omega)&=&
    \int_{E_F}^{+\infty} n_{d -\sigma}(\epsilon) [U_{cv}- T_{hh}^{cv}(\omega\!-\!\epsilon)\cdot \nonumber\\
    &&\qquad\qquad\qquad\cdot(1+U_{cv}A_{cd}(\omega\!-\!\epsilon))] d\epsilon \nonumber\\
    &+& \int_{E_F}^{+\infty} n_{d \sigma}(\epsilon) [(U_{cv}-J_{cd}) -\tilde{T}_{hh}^{cv}(\omega\!-\!\epsilon) \cdot \nonumber\\
    &&\qquad \qquad \cdot [1+(U_{cv}\!-\!J_{cv})] \tilde{A}_{cd}(\omega\!-\!\epsilon) d\epsilon.
\end{eqnarray}
Here $n_{d \sigma}(\epsilon)$ is the $d$ contribution to the valence density of states of spin $\sigma$.   The hole-hole scattering associated to the on-site interactions is described by T-matrices $T_{hh}^{cv}$ and $\tilde{T}_{hh}^{cv}$ for scattering between antiparallel and parallel spin holes respectively; similarly $A_{\sigma}$ ($\widetilde{A_{\sigma}}$) includes antiparallel (parallel) electron-hole scattering \cite{rozzi2000}.

\begin{figure}[h!]
 \begin{center}
\includegraphics[width=9cm]{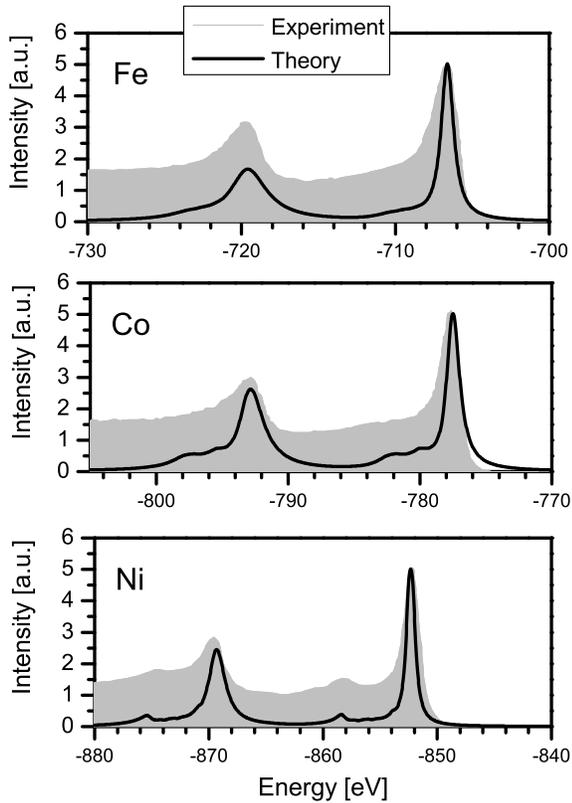}
 \bigskip
  \caption{\label{coresf} 2P core level photoemission for Fe, Co and Ni
inclusive of correlation effects (black line) compared with
experimental ones \cite{coreexp} (grey area). }
\end{center}
\end{figure}

Fig. \ref{coresf} shows the calculated spectral functions for the 2$p$ core level of Fe, Co and Ni obtained assuming the $U$ and $J$ values of table \ref{UJ} with binding energies adjusted to the experimental values \cite{coreexp}. Since our calculated self-energy  is responsible of an intrinsic broadening  identical at the two edges, to reproduce the observed different lifetimes at the two edges  we have added  extra broadening  $\delta=0.5 eV$ and , $\delta=1.5 eV$ for  $2p_3/2$ and $2p_2/2$  respectively. The spectral functions reproduce the essential features of the core photoemission data for transition metals where a main peak is always followed by a satellite structure at higher binding energies. This structure is the fingerprint of e-e correlation being commonly attributed to the admixture of  different single particle configurations: not just the ground state with one core electron
missing but also the shake-up states where electrons are promoted to
higher energy levels.  These physical processes are contained in our
many body description based, as previously mentioned, on an
extension to the solid state of the configuration interaction
scheme.

We turn now to  valence states. The self-energy correction to a
valence band level  $\epsilon_{k n \sigma}$ is slightly more complicated
with respect to the core one reflecting the  complex valence band
structure where different orbitals are inextricably hybridized (for
a detailed description of the valence self-energy see reference \cite{Bellini}).

\begin{figure}
\includegraphics[width=9cm]{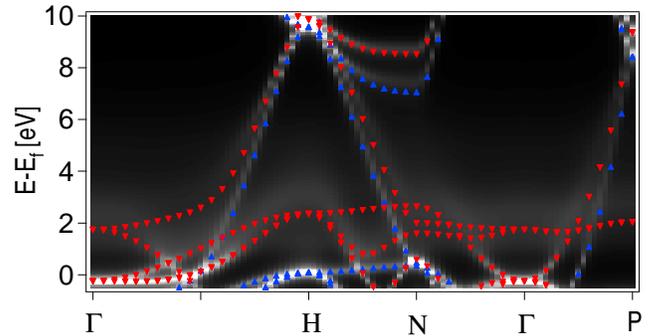}
  \caption{(color online)\label{valsf} Iron empty states: interacting spectral functions are shown as color map and compared with single particle eigenstates reported as red (spin down) and blu (spin up) arrows. In this energy region self-energy
corrections for majority spin states are negligible.}
\end{figure}

The effect of correlation on empty states is shown in figure \ref{valsf} for iron where the electron spectral
functions are shown along high symmetry directions of the Brillouin
zone. Notice that only some of the empty bands (the minority spin
ones of $d$ orbital character) are  affected by self-energy
corrections, mainly in terms of life time broadening. The overall
effect of correlation on empty states is however very small, and
even smaller for Cobalt and Nickel where $d$ states are almost fully
occupied.

The response of interacting electrons to the creation of an e-h pair
due to X-ray absorption is  connected to these hole and particle
spectral functions:  within the single particle approach  the
absorption cross section is just the joint density of core and
valence states modulated by the dipole matrix elements
\begin{equation}
\label{mu0}
 \mu^{\pm}_0(\omega) \propto \sum_{kn} |D^{\pm}_{ckn}|^{2}\sum_{\sigma}Im \chi^0_{c kn \sigma}(\omega)
\end{equation}
with
\[
Im \chi^0_{c kn \sigma}(\omega) =
\int_{-\infty}^{\infty} \delta (\Omega-\epsilon_{c\sigma}) \delta
(\omega+\Omega-\epsilon_{kn \sigma}) d \Omega
\]

One effect of many body correlation is to replace  this joint density of
states  with the convolution of hole and particle  spectral functions. Formally
this corresponds to calculate a two-particle polarization propagator
$\chi_{c v}(\omega)$ where the  bare particle and hole propagators
are substituted with the dressed ones
\begin{equation}
\label{mu}
 \mu^{\pm}(\omega) \propto \sum_{kn} |D^{\pm}_{ckn}|^{2}\sum_{\sigma}Im \chi_{c kn \sigma}(\omega)
\end{equation}
with
\[
\label{imchi} Im \chi_{c kn \sigma}(\omega) = \int
A^c_{\sigma }(\Omega) A_{kn\sigma}(\Omega+\omega) d \Omega
\]

 $D^{\pm}_{ckn}$ appearing both in eq. (\ref{mu0}) and eq. (\ref{mu}) is the matrix element of the electric dipole moment over the core
and valence states

\begin{equation}
 \label{eqc1_1}
D^{\pm}_{ckn}= \sum_{m_j=-j}^j\langle \Psi^c_{j m_j} | \hat{
\epsilon}^{\pm} \cdot e \textbf{ r }| \Psi_{kn} \rangle
\end{equation}
with $\hat{ \epsilon}^{\pm}$  the polarization
vector (left/right) of the incident photon, $e$ the electron charge, $\Psi^c_{j m_j}$ and $\Psi_{kn}$ the spin-orbit split core and valence eigenstates respectively..
The XMCD signal is obtained as the difference $ \mu^{+}(\omega)- \mu^{-}(\omega)  $.

\begin{figure}
\includegraphics[width=9cm]{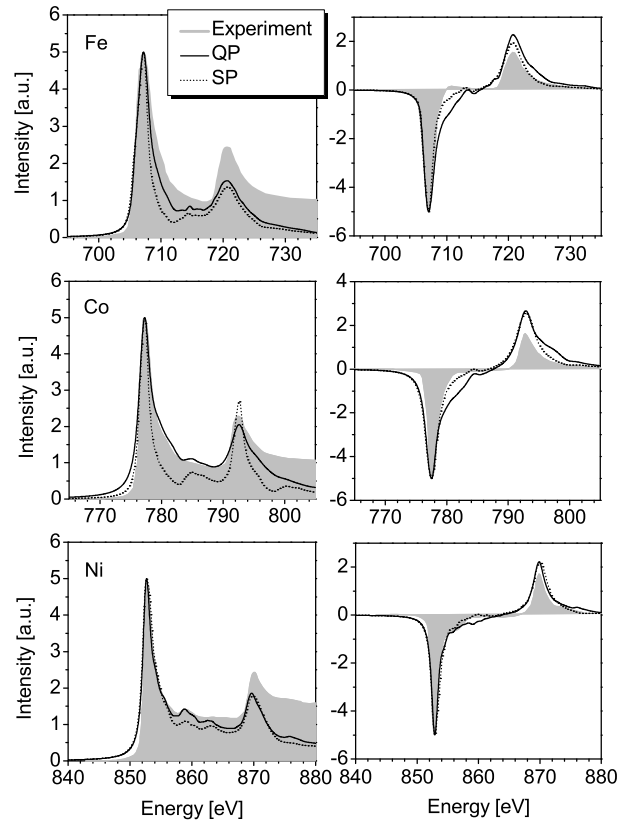}
  \caption{\label{fig3} Absorption (left panels) and dichroic(right
panels) spectra of Fe, Co, Ni. Theoretical spectra, with (QP) and without (SP)
self-energy corrections are compared with experimental ones
\cite{chen1994,chen1990} represented by  gray areas.}
\end{figure}

Figure \ref{fig3}  shows the calculated absorption (a) and dichroic (b) spectra for Fe, Co and Ni calculated both with and without self-energy corrections, compared with experimental data \cite{chen1990,chen1994}. We notice that in most cases the inclusion of many-body effects improves the agreement with experimental XAS spectra, making the lines   more asymmetric. What makes line shapes asymmetric is the presence of shake-up satellite structures in the hole-spectral functions. These structures are
essential to recover asymmetric line shapes also in XMCD spectra.
Since in our calculation the $L_2$ and $L_3$ lines have  the same line
shape by construction, the only difference being their width, it is not surprising
that the asymmetry in the dichroic lines is the same at the two edges. In the case of nickel  where both peaks in the measured dichroic spectrum present the same pronounced asymmetry,  the inclusion of e-e correlation improves the agreement between theory and experiment. For iron and cobalt on the contrary we are able  to reproduce rather well the dichroic line shape at the $L_2$ edge but less accurately the  $L_3$  one which turns out to be experimentally rather symmetric and more similar to the uncorrelated result. The asymmetry of the $L_2$ edges increases from Fe to Ni, in agreement with the corresponding trend in the core-valence electronic correlation described by $U_{pd}$ (see Tab. 1).

As we mentioned at the beginning what is still missing in this description is the e-h attraction
responsible of  excitonic effects. These effects can be included by considering the two-particle
eigenvalue problem assuming the excited states of the N-particle interacting system
to be a superposition of single particle states with one core hole and an electron above the Fermi level.
This is the so-called Tamm-Dancoff approximation (TDA) \cite{Rohlfing} and is at the core also of the
3BS scheme applied here to describe the charged excitations - just one e-h pair added in this
case to the $N\pm 1$ -particle interacting system. TDA has been used to solve the Bethe-Salpeter equation (BSE)
and treat excitonic effects in valence-valence transitions \cite{Onida} and more recently extended
to X-ray absorption from core states \cite{Rehr1,Ambrosch-Draxl}. In both cases
the dressed one-particle propagators have been obtained within the GW approximation \cite{GW}.
In the present case we are interested in a regime of high e-e correlation where the a
perturbative approach like GW is supposed to be less appropriate being unable to reproduce
satellite structures that dominate the core level spectra.
A non-perturbative method such as the 3BS approach would be more appropriate and
we are presently testing the extension of our theory in this direction.

A further difference between the present approach and the  ab-initio methods based
on either GW+BSE \cite{Rehr1,Ambrosch-Draxl} or time-dependent local density approximation
\cite{Ankudinov} is the treatment of the e-e interaction: in these schemes the electrons
interact through the full long range,  dynamically screened Coulomb potential while in
our approach only the short-range part of the Coulomb interaction is considered,
in the spirit of the Hubbard model. The implicit assumption, common to most approaches
used to describe e-e correlation in strongly correlated materials,  is to attribute many body
effects to just the on-site interaction, the long range part being included at the mean-field level
in the  band calculation. This assumption  seems particularly reasonable in the case of the
strongly localized core states but in practice leads to  the introduction of adjustable parameters.
This is has been up to now a shortcoming of most the approaches based on the Hubbard model, with the advantage to allow for solutions beyond the perturbation limit.

In summary, we have shown that many-body effects due to e-e interactions can be included
in the description of  X-ray absorption and dichroic spectra by replacing the joint density
of states appearing in the absorption cross-section with a convolution of (core) hole and
(valence) electron spectral functions.  Whenever e-e correlations modify significantly hole
and electron spectral functions we expect non negligible effects in the absorption spectra as well.
In the case of 3$d$ transition metals these modifications are  essentially associated to the
shake-up structures that appear in the core-level photoemission spectra as a residue of atomic multiplets.
These structures,  related to the admixture of  different single particle configurations,
can be reproduced only in a many body approach and are responsible  of the  asymmetric
line shapes of XAS and XMCD structrues at L$_{2,3}$ edges in Fe, Co and Ni.
In other systems where local e-e correlations  affect significantly the empty part of the
valence band  we expect even more drastic changes: this should be the case  for instance
of highly correlated materials  where on site e-e repulsion is responsible of the creation of extra empty (Hubbard) bands \cite{silke}.

Finally we want to stress that our approach has  a purely experimental counterpart:
it would be interesting to check on an  experimental basis wether - and to what extent -
the measured absorption spectra can be reproduced as a straightforward  convolution of
direct and inverse  photoemission  data, in this way confirming the direct link between neutral excitations,
due to X-ray absorption, and charged excitations corresponding to  hole and electron addition,
link that is clear theoretically  but that has not been demonstrated experimentally so far.

\begin{acknowledgments}
This research was supported in part by the National Science Foundation under Grant No. PHY05-51164 and  by MIUR - Prin 2008.
\end{acknowledgments}


\end{document}